\documentclass[useAMS,usenatbib]{mn2e}

\usepackage{aas_macros} 
\usepackage{graphicx} 
\usepackage{tabularx} 
\usepackage{multirow} 
\usepackage{hhline} 
\usepackage{subfigure} 
\usepackage{amssymb} 
\usepackage{flafter} 
\usepackage{psfrag} 

\newcommand{\eg}{e.g. }

\title[The Arcminute Microkelvin Imager]{The Arcminute
Microkelvin Imager}
\author[Zwart et al.]{AMI Consortium: 
J.~T.~L.~Zwart\thanks{Issuing author -- e-mail: jtlz2@mrao.cam.ac.uk},
R.~W.~Barker,
P.~Biddulph,
D.~Bly,\newauthor
R.~C.~Boysen,
A.~R.~Brown,
C.~Clementson,
M.~Crofts,
T.~L.~Culverhouse$^{1}$,\newauthor
J.~Czeres,
R.~J.~Dace,
M.~L.~Davies,
R.~D'Alessandro,
P.~Doherty,
K.~Duggan,\newauthor
J.~A.~Ely,
M.~Felvus,
F.~Feroz,
W.~Flynn$^{2}$,
T.~M.~O.~Franzen,
J.~Geisb\"usch,\newauthor
R.~G\'enova-Santos$^{3}$,
K.~J.~B.~Grainge,
W.~F.~Grainger$^{4}$,
D.~Hammett,\newauthor
R.~E.~Hills$^{5}$,
M.~P.~Hobson,
C.~M.~Holler$^{6}$,
N.~Hurley-Walker,
R.~Jilley,\newauthor
M.~E.~Jones$^{6}$,
T.~Kaneko,
R.~Kneissl$^{7}$,
K.~Lancaster$^{8}$,
A.~N.~Lasenby,\newauthor
P.~J.~Marshall$^{9}$,
F.~Newton,
O.~Norris$^{2}$,
I.~Northrop,
D.~M.~Odell,
J.~C.~Pober,\newauthor
G.~G.~Pooley,
V.~Quy,
C.~Rodr\'{\i}guez-Gonz\'{a}lvez,
R.~D.~E.~Saunders,
A.~M.~Scaife,\newauthor
J.~Schofield,
P.~F.~Scott,
C.~Shaw,
T.~W.~Shimwell,
H.~Smith,
A.~C.~Taylor$^{6}$,\newauthor
D.~J.~Titterington,
M.~Veli\'{c}$^{10}$,
E.~M.~Waldram,
S.~West,
B.~A.~Wood
and G.~Yassin$^{6}$\\
Astrophysics Group, Cavendish Laboratory, J.~J.~Thomson Avenue, Cambridge, CB3 0HE\\
Present addresses:\\
$^{1}$Kavli Institute for Cosmological Physics, University of Chicago, 5640 South Ellis Avenue, Chicago, IL 60637, USA\\
$^{2}$The University Chemical Laboratory, Lensfield Road, Cambridge, CB2 1EW\\
$^{3}$Instituto de Astrof\'isica de Canarias, 38200 La Laguna, Tenerife, Canary Islands, Spain\\
$^{4}$Columbia University, Department of Physics, 538 W. 120th Street, New York, NY 10027, USA\\
$^{5}$Joint ALMA Office, Av El Golf 40, Piso 18, Santiago, Chile\\
$^{6}$Astrophysics group, Department of Physics, Denys Wilkinson Building, Keble Road, Oxford, OX1 3RH\\
$^{7}$MPI f\"ur Radioastronomie, Auf dem H\"ugel 69, D--53121 Bonn, Germany\\
$^{8}$Astrophysics Group, H.~H.~Wills Physics Laboratory, Royal Fort, Tyndall Avenue, Bristol, BS8 1TL\\
$^{9}$UC Santa Barbara, Santa Barbara CA 93106, USA\\
$^{10}$Spatial Technology Ltd, 1 Quayside, Cambridge, CB5 8AB
}
\begin{document}

\date{Accepted ????; received ????; in original form ????.}

\pagerange{\pageref{firstpage}--\pageref{lastpage}} \pubyear{2008}

\maketitle

\label{firstpage}


\begin{abstract}
The Arcminute Microkelvin Imager is a pair of interferometer arrays
operating with six frequency channels spanning $13.9$--$18.2$~GHz,
with very high sensitivity to angular scales
$30\arcsec$--$10\arcmin$. The telescope is aimed principally at
Sunyaev-Zel'dovich imaging of clusters of galaxies. We discuss the
design of the telescope and describe and explain its electronic and
mechanical systems.
\end{abstract}


\section{Introduction}
\label{sec:intro}

Clusters of galaxies are the best samplers of matter on the largest
scales and are sensitive probes of structure formation, both in the
linear regime of growth and when merging, shocking and gradual
virialization occur. Determining the structures and physics of
clusters, their mass function and their evolution is therefore of
basic importance.

Observationally, however, relatively little is known about the
properties of the cluster population at redshifts $z\gtrsim 1$, due to
the fundamental problem of the dimming of surface brightness with
redshift. Optical observations suffer from confusion from foreground
galaxies, and both optical and X-ray observations suffer from biases
towards strong concentrations of mass.

Ionized matter at any redshift out to recombination will
inverse-Compton scatter the cosmic microwave background (CMB)
radiation and so imprint spatial fluctuations upon it. Hot ionized gas
in the deep potential wells of galaxy clusters (statistically)
up-scatters CMB photons passing through it, giving rise to the
Sunyaev-Zel'dovich (SZ) effect (\citealt{sz70}, \citealt{sz72}; see
\eg \citealt{birkinshaw99} and \citealt*{carlstrom02} for reviews); at
frequencies below 217~GHz, one sees a decrease in the CMB temperature
towards the cluster. This dip in temperature $\Delta T$ is given in
the Rayleigh-Jeans region by $\Delta T = -2yT_{\rm{CMB}}$, where the
Comptonization parameter $y=\int n_{\rm{e}} \sigma_{\rm{T}}
\frac{k_{\rm{B}}T_{\rm{e}}}{m_{\rm{e}} c^2} {\rm d}l$,
$\sigma_{\rm{T}}$ is the Thomson cross-section, $l$ is the
line-of-sight path, and $m_{\rm{e}}$, $n_{\rm{e}}$ and $T_{\rm{e}}$
are respectively the electron mass, density and temperature. The dip
$\Delta T$ is thus proportional to the line integral of pressure but
does \textit{not} depend on redshift $z$. One way of seeing this is
that in such a inverse-Compton scattering process, the power lost by
the electrons and given to the CMB photons is proportional to the
energy density of the CMB radiation which is proportional to
$(1+z)^{4}$; this exactly cancels out the cosmological drop in
bolometric surface brightness proportional to $(1+z)^{-4}$. This is an
extremely important result --- one can get \textit{directly} to high
redshift using SZ without any intermediate steps.

The total SZ decrement in flux density, $ \Delta S_{\rm SZ}$, is
proportional to the integral of the brightness temperature $ \int
\Delta T_{\rm SZ}\,{\rm d}\Omega$ over the solid angle $\Omega$
subtended by the cluster. Thus $\Delta S_{\rm{SZ}} \propto
D_{\rm{A}}^{-2} \int n_{\rm{e}}T_{\rm{e}} \, \rm{d}V, $ where
$D_{\rm{A}}$ is the angular-size distance and $\rm{d}V$ is an element
of volume. For concordance cosmologies, for redshifts between, say,
0.3 and 3, $D_{\rm{A}}$ is only weakly dependent on redshift. Since
$D_{\rm{A}}$ is always a weaker function of redshift than luminosity
distance $D_{\rm{L}}$, SZ clusters will have a selection function that
is much less dependent on redshift then most self-luminous
objects. $\Delta S_{\rm{SZ}}$ measures the total thermal energy of the
cluster, and since temperature is strongly correlated to mass (\eg $T
\propto M^{2/3}$ for a population of clusters, modelled as virialized
gravitationally-collapsed systems), the integrated SZ flux density
measures gas mass directly, and without bias to concentrated
structure. Detailed studies of clusters in SZ and other wavebands will
help to establish the cosmological relationship between $M$ and $T$
and the thermal history of the cluster, and will help to calibrate the
cluster scaling relations.

These points immediately emphasize the importance of SZ surveys for
clusters. These have long been advocated (see \eg \citealt*{ksy86},
\citealt{bm92}, \citealt{bs94}, et seq.) and it is now widely
recognized that such surveys are key measurements for cosmology. For
example, it is critical to measure and understand the comoving number
density of clusters, $N(M,z)$, as a function of mass $M$ and
redshift. $N(M,z)$ is a very strong function of the key cosmological
parameter $\sigma_8$ --- the density contrast on scales of
$8\,h^{-1}$Mpc now --- and of the physics of cluster assembly.

The X-ray flux from a cluster is proportional to
$n_{\rm{e}}^{2}f(T_{\rm{e}})$, where $f(T_{\rm{e}})$ is not a strong
function of $T_{\rm{e}}$. The SZ flux is proportional to $n_{\rm{e}}
T_{\rm{e}}$, so the combination of good X-ray and SZ data on a cluster
gives robust determinations of density and temperature distributions
and the related physics. Since the SZ flux density is proportional to
the integrated gas pressure rather than to the density squared, it is
relatively more sensitive to the outer parts of clusters than are
X-rays, and not biased towards dense regions.

All these points also emphasize the importance of pointed,
high-resolution SZ observations of clusters known from X-ray,
optical/IR or SZ surveying, as well as X-ray/optical/IR observations
of SZ-selected targets. For example, cluster observations from
XMM-Newton and Chandra show clear entropy structures in a cluster's
gas, with contact discontinuities and sloshing motions (see \eg
\citealt{rossetti07}), and bubbling and reheating of cooling gas by
FRI radio jets of low luminosity but considerable bulk kinetic power
\citep[\eg ][]{sanders07}. Observations in different wavebands tease
out different selection effects and allow for cross-calibration.

The first unequivocal SZ detections were made using the OVRO 40-m
telescope and the NRAO 140-foot telescope
\citep*{birkinshaw81,birkinshaw-plus84,uson86}, and the first SZ image
was obtained with the Ryle Telescope (RT; \citealt{jones93}). The RT
and OVRO/BIMA interferometers have since then made many SZ
observations of known clusters (see \eg \citealt{grainge96},
\citeauthor{grainge02a} \citeyear{grainge02b}a,b,
\citeauthor{cotter02a} \citeyear{cotter02b}a,b, \citealt{grainger02},
\citealt{saunders03}, \citealt{jones05}; \citealt*{carlstrom96},
\citealt{grego00}, \citealt{patel00}, \citealt{reese00},
\citealt{joy01}, \citealt{reese02}, \citealt{bonamente06},
\citealt{laroque06}). Each of these observations has however been long
(1--60 days), because even the shortest baselines resolved out most of
the SZ signal: some 90 per cent of the SZ flux from a massive cluster
at $z=0.2$ is missed by a $600$-$\lambda$ baseline, the smallest
possible given the existing dish sizes of the RT and OVRO/BIMA.

Several other instruments have successfully observed the SZ effect,
including the OVRO 5.5-m Telescope \citep{herbig95}, SuZIE
\citep{holzapfel97}, ACBAR \citep{cantalupo02}, CBI
\citep{udomprasert04}, VSA \citep{lancaster05} and OCRA
\citep{lancaster07}. It has, however, long been clear that, for
detailed imaging and blind surveying, a new generation of fast SZ
imagers is required, with angular-scale and brightness-temperature
sensitivities far better matched to the features of interest in
clusters. Of course, such instruments should also be able to measure
the CMB power spectrum at high angular scales ($\ell > 2000$) and thus
investigate the origins of, for example, the excess CMB power detected
by CBI \citep{cbi-excess}. They also are pertinent to the search for
other non-Gaussian features such as topological defects.

Accordingly, we have designed, built and now operate the Arcminute
Microkelvin Imager (AMI, see \citealt{rk01} and AMI Collaboration
\citeyear{ami-a1914}), a pair of interferometer arrays operating
around 15~GHz near Cambridge, UK. Before describing AMI in detail, we
next review other new SZ instruments, as well as the advantages and
disadvantages of interferometers.

\subsection{Fast SZ Instruments}
\label{sec:intro:instruments}

Several powerful SZ instruments are now coming online. The majority
are direct-imaging, focal-plane arrays of bolometers, operating at
frequencies above 90~GHz: ACT (\eg \citealt{ACT06}) in Chile uses a
6-m antenna with a resolution of $1\farcm44$ at \eg 150~GHz; APEX (\eg
\citealt{APEX06}) in Chile has a 12-m antenna with $1\farcm0$
resolution at \eg 150~GHz; SPT (\eg \citealt{SPT04}) at the South Pole
has a 10-m antenna with resolution of $1\farcm06$ at \eg 150~GHz; the
Planck satellite, due for launch in 2009 January, will detect clusters
in SZ over the whole sky, but detecting distant clusters will be
challenging due to its limited resolution (see \eg
\citealt*{geisbuesch05}).

There are also three interferometers, operating at lower frequencies:
AMI, see below; AMiBA (see \eg \citealt{amiba06}) in Hawaii, with the
array comounted on a hexapod and currently at 90~GHz; and the SZA (see
\eg \citealt{sza07}) in California with 3.5-m diameter antennas ---
the SZA has been operating at 30~GHz with a resolution of $1\arcmin$
and will be operated at 90~GHz as part of CARMA
(\texttt{www.mmarray.org}).

\subsection{Advantages and Disadvantages of Interferometers}
\label{sec:intro:ints}

Interferometry has some significant advantages for observations where
high sensitivity and low systematics are required (see \eg
\citealt{church95}, \citealt{lay00}, \citealt{katy-thesis} and
\citealt{zwart2007}):

\begin{enumerate}

\item Stability of receivers. Short-term fluctuations in the gains of
the front-end amplifiers are uncorrelated and therefore contribute
only to the random noise.

\item Emission from the atmosphere is largely uncorrelated (provided
the antenna beams do not overlap within the troposphere), and
associated receiver power fluctuations are completely uncorrelated and
thus contribute only to the system noise.

\item Fringe-rate filtering. Astronomical signals are modulated at the
celestial fringe rate by the Earth's rotation. Unwanted interference
from sources at terrestrial fringe rates, such as ground spill and
geostationary satellite broadcasts, can be attenuated. Residual
correlated atmospheric emission also has very little power in
components that are synchronous with the astronomical fringe rate.

\item  Interferometers have zero response to power which is uniformly 
spatially distributed (\eg most of the atmospheric emission and the CMB 
average temperature), and hence do not respond to temporal 
fluctuations in that power.

\item Each interferometer baseline responds to only a narrow range of
spatial frequencies; as well as filtering out the large-scale
atmospheric signals this can be used to \eg minimize the response to
the primary CMB anisotropies while retaining sensitivity to clusters.

\item Foreground removal. Radio point sources can be detected by
baselines chosen to be sensitive to smaller angular scales. This
allows them to be measured simultaneously and subtracted.

\end{enumerate}

\noindent The main disadvantages of interferometers are complication
(with the number of correlators going as the square of the number of
detector elements) and a restriction on observing frequency. The SZ
decrement has a maximum amplitude in intensity at approximately
130~GHz; contaminating synchroton emission from radio sources
\textit{generally} decreases with increasing frequency (falling in
temperature as $\nu^{-(\alpha + 2)}$ where $\alpha$ is the
flux-density spectral index), although the effect of dust in galaxies
on SZ observation at, say, 150~GHz is presently uncertain. Radio
interferometers employing heterodyne receivers have costs and
amplifier noise that rise with frequency, which in the past has
limited interferometers to observing frequencies of around 30~GHz.

\subsection{The Rest of this Paper}
\label{sec:intro:rest}

We next describe the design of the instrument (section
\ref{sec:design}), including the choice of frequency and the need for
two interferometric arrays, the Large Array (LA) and the Small Array
(SA). We describe the antennas, optics and array configurations of the
SA and LA in section \ref{sec:ant}. The amplifiers and cryostats are
outlined in section \ref{sec:cryo}, with the signal chain expounded in
section \ref{sec:if}. Section \ref{sec:corr} covers correlator and
readout systems, and section \ref{sec:control}, the telescope control
systems. Some lessons learned are outlined in section \ref{sec:comm}.

\begin{table}
\centering
\caption{AMI technical summary.}\label{table:summary}
\begin{tabular}{{l}{c}{c}}
\hline
                           & SA                 & LA                  \\
\hline
Antenna diameter           & 3.7~m              & 12.8~m              \\
Antenna efficiency         & 0.75               & 0.67                \\
Number of antennas         & 10                 & 8                   \\
Number of baselines        & 45                 & 28                  \\
Baseline lengths (current) & 5--20~m            & 18--110~m           \\ 
Primary beam (15.7~GHz)    & $20\farcm1$        & $5\farcm5$          \\
Synthesized beam           & $\approx3\arcmin $ & $\approx30\arcsec$  \\
Flux sensitivity           & 30~mJy~s$^{-1/2}$  & 3~mJy~s$^{-1/2}$    \\
Observing frequency        & \multicolumn{2}{c}{13.9--18.2{~GHz}}     \\
Bandwidth                  & \multicolumn{2}{c}{4.3{~GHz}}            \\
Number of channels         & \multicolumn{2}{c}{6}                    \\
Channel bandwidth          & \multicolumn{2}{c}{0.72{~GHz}}           \\
System temperature         & \multicolumn{2}{c}{25~K}                 \\
Declination range          & $>-15\degr$        & $>-20\degr$         \\
Elevation limit            & $+20\degr$         & $+5\degr$           \\
Polarization measured      & \multicolumn{2}{c}{I+Q}                  \\
\hline
\end{tabular}
\end{table}

\section{AMI Basic Design}
\label{sec:design}

\subsection{Choice of Frequency}
\label{sec:design:freq}

\begin{figure}
\begin{center}
\includegraphics[width=7cm,origin=br,angle=0]{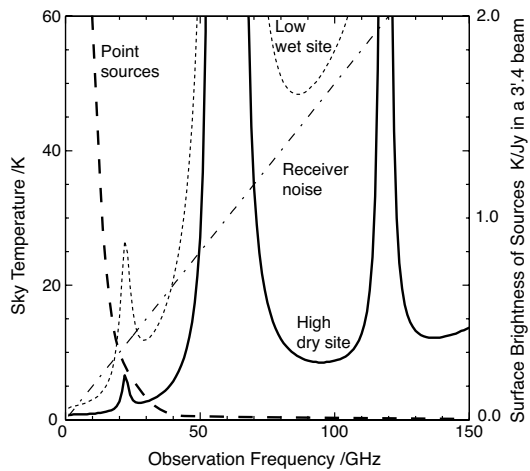}
\end{center}
\caption{The contribution of atmospheric emission to system
temperature as a function of frequency. The upper dashed curve is for
a sea-level site like Cambridge. The lower solid line is for a high,
dry site like Mauna Kea. The typical noise temperature of a HEMT
amplifier is also plotted. The thick dashed curve is the brightness
temperature of an unresolved 1-Jy source (with a spectral index of
0.7) at 15~GHz in a $3\farcm4$ beam. (From \citealt{kaneko2006}, after
\citealt{jones02}).\label{fig:temp2}}
\end{figure}

The AMI frequencies are chosen to give the highest ratio of SZ signal
to total noise, constrained by our need to minimize cost. The
competing noise terms are the receiver noise and sky background, which
rise with frequency, and the effect of radio source confusion, which
falls with frequency. The temperature of the atmosphere at a
low-altitude site rises quite steeply between $\approx 5$~K at 15~GHz
and $\approx 30$~K at 30~GHz (see Figure \ref{fig:temp2}). Experience
from the 5-km Telescope \citep{ryle72}, CAT \citep{robson94}, and the
initial testing of the VSA at 30~GHz at the Lord's Bridge site near
Cambridge, UK \citep{rusholme2001}, showed that observations at
frequencies up to 18~GHz are achievable from there, but sensitive
30-GHz observations are ruled out. However, the existence of the Ryle
Telescope (which has eight 13-m antennas) on this site means that high
sensitivity and resolution observations of the confusing point sources
are possible. Constructing a higher-frequency instrument on a remote
site (there is none suitable in the UK) would have been very much more
expensive. Hence we chose to operate in the band 12--18~GHz, the upper
limit being set by the presence of the 22-GHz water line, and the
bandwidth set by the use of single-moded waveguide and feeds. Note
that the precise frequency range and bandwidth are dependent on the
dimensions of commercially available waveguide.

\subsection{The Two Arrays}
\label{sec:design:dual}

The choice of such frequencies for SZ observation leads immediately to
the requirement to provide much higher flux sensitivity on longer
baselines to remove the small-angular size radio sources which would
otherwise obscure the SZ (see \eg \citealt{grainge96},
\citealt{grainger02}, \citealt{lin08}). The eight, 13-m antennas of
the RT --- the RT itself being a radical receiver- and
signal-processing upgrade of the 5-km Telescope --- form the basis of
source removal. We thus arrived at a two-array design, with the SA
sensitive to SZ and contaminant sources, and the LA for robust removal
of the sources. Note, however, that the huge flux sensitivity of the
LA and its particular range of baselines combine to allow it also to
map SZ at high resolution.

\begin{figure}
\begin{center}
\psfrag{ell}{$\ell$}
\includegraphics[width=8cm,origin=br,angle=0]{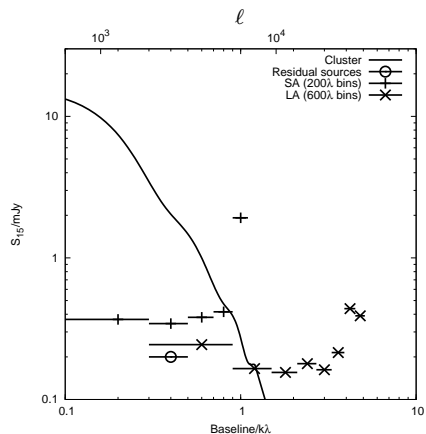}
\end{center}
\caption{This figure illustrates of some of the issues affecting the
sensitivity of AMI to an SZ signal. The solid line indicates the SZ
signal received by the SA from a typical rich cluster (in this case
A2218 at $z=0.171$). The vertical crosses indicate the 1-$\sigma$ flux
sensitivity of the SA in $200$-$\lambda$ bins for an 8-hour
observation; diagonal crosses show the 1-$\sigma$ flux sensitivity of
the LA (with $600$-$\lambda$ bins) for an observation of the same
duration and rastered in order to cover the same sky area; for both
the SA and the LA, the horizontal lines denote the range of baselines
contributing to each cross. The circle indicates the confusion from
remnant unsubtracted sources in a $2\arcmin$ FWHM synthesized beam,
with source count from \citealt{waldram03}, and assuming that all
sources above a limiting flux density $S_{\mathrm{lim}}=1.9$~mJy have
been removed; the LA can measure the fluxes of each of these sources
at $\ge 5\sigma$ in about 15~mins over an SA
field-of-view. Contamination by priomordial CMB signals is not
significant for AMI observations of clusters.
\label{fig:grainge}}
\end{figure}

The details of the SA and LA are outlined in Table \ref{table:summary}
and sensitivity illustrated in Figure \ref{fig:grainge}. Note that
maintaining full angular resolution over the field of view of each of
these arrays would be impossible with a single 6-GHz-wide channel
because of chromatic aberration. Hence we split the signal into eight
0.75-GHz-wide channels, which results in acceptably small chromatic
aberration at the field edge at full array resolution.

\begin{figure}
\begin{center}
\includegraphics[width=7cm,origin=br,angle=0]{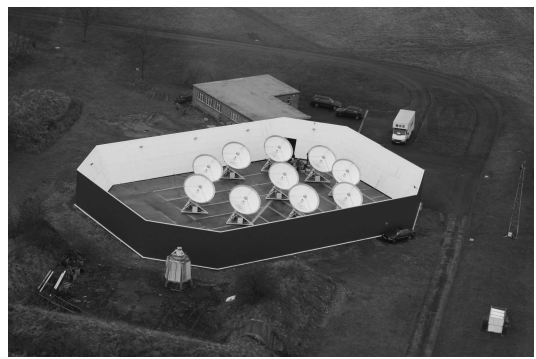}
\end{center}
\caption{SA from the south east. The correlator is in the building
behind. The enclosure floor is now covered in
aluminium.\label{fig:ami-sa}}
\end{figure}

\begin{figure}
\begin{center}
\includegraphics[width=7cm,origin=br,angle=0]{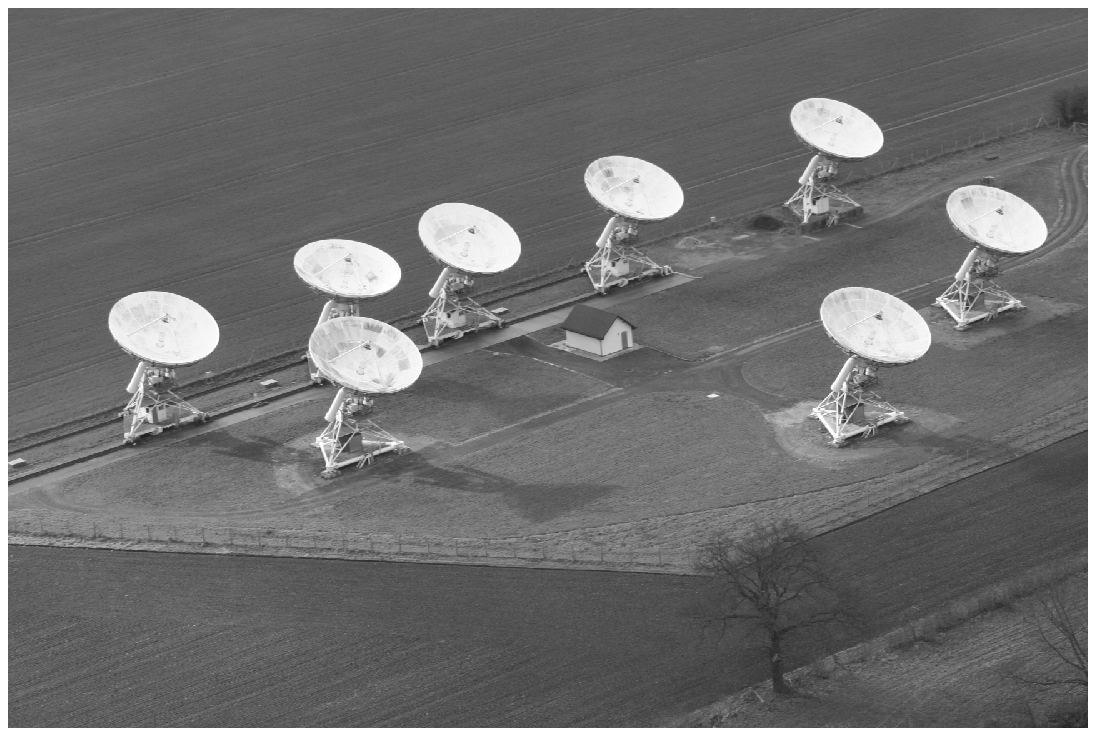}
\end{center}
\caption{LA from the north east. The correlator is in the central
hut. The railway track, which runs diagonally across the photograph,
is approximately east-west. Four antennas are moveable along the
track. \label{fig:ami-la}}
\end{figure}

\section{Antennas and Optics}
\label{sec:ant}

\begin{figure*} 
\begin{center}
\includegraphics[width=15cm,origin=br,angle=0]{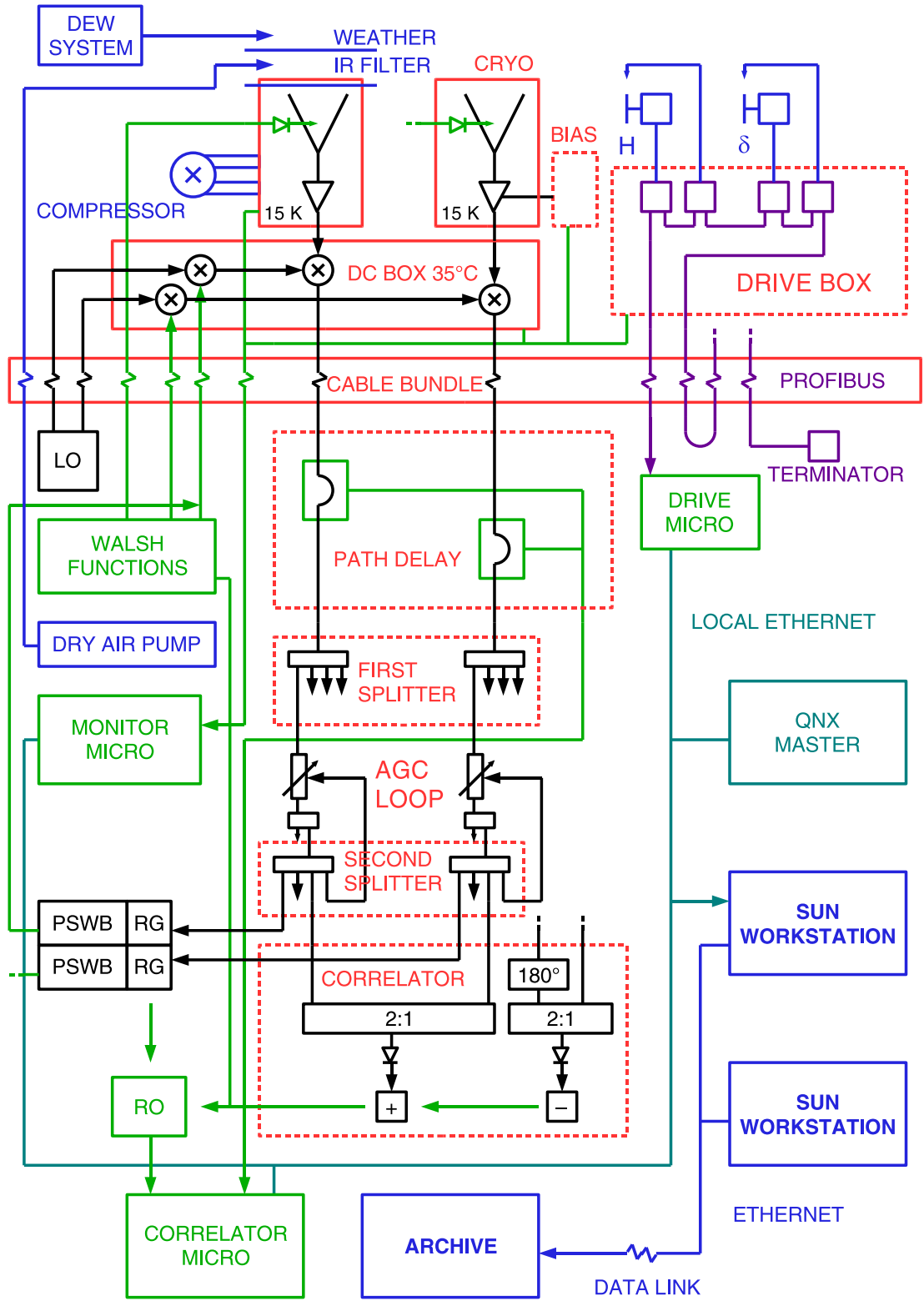}
\end{center}
\caption{SA sub-systems. Equalizers and all amplifiers (apart from the
front-end amplifiers) are omitted from this diagram. AGC is automatic
gain control, DC is downconversion, LO is the local oscillator, PSWB
is phase-switch balance, RG is the system temperature monitor (`rain
gauge') and RO is correlator readout. The LA sub-systems are
essentially identical (see text for details), but in particular: there
is no dew system, dry-air pump or infra-red filter, in part because
the horn extends beyond the cryostat; the drive system does not use
the Profibus interface, with instead one drive microprocessor per
antenna; monitoring is carried out by each antenna's drive
microprocessor; and there is no QNX system.\label{fig:sa-system}}
\end{figure*}

The SA and LA are shown in Figures \ref{fig:ami-sa} and
\ref{fig:ami-la}. The SA sub-systems are illustrated in Figure
\ref{fig:sa-system} (the LA sub-systems are very similar, but see
section \ref{sec:if:differences}).

\subsection{Small Array (SA)}
\label{sec:ant:sa}

The SA consists of 10 3.7-m diameter antennas surrounded by an
aluminium groundshield. These are described in the following
subsections.

\subsubsection{Enclosure and Mount}
\label{sec:ant:sa:mount}

\begin{figure*}
\centering
\includegraphics[width=12cm]{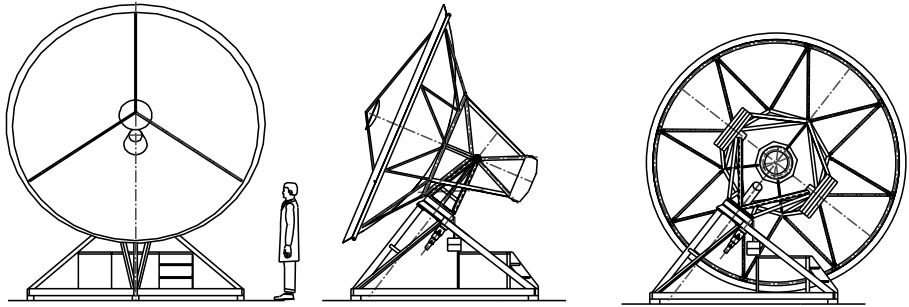}
\caption{A single SA antenna.\label{fig:sa-antenna}}
\end{figure*}

The antennas of the SA are designed around commercially-available spun
3.7-m parabolic reflectors. These are made by fixing an aluminium
sheet to the centre of a paraboloidal former; the whole assembly is
then spun about its symmetry axis and the sheet is pressed onto the
former from the centre outwards with a roller. The rms surface
accuracy of these single-piece reflectors is better than 0.3~mm and
the cost savings over machined or segmeneted dishes are large. Like
the existing LA dishes, the SA dishes are equatorially mounted with a
Cassegrain focus (Figure \ref{fig:sa-antenna}). An advantage of an
equatorial mount is that the observed angle of polarization is
constant over the whole observation. Point sources measured by the LA
can then be directly subtracted from the SA data.

Each dish has a circular tube rolled into the rim to provide
stiffness. The backing structure consists of a lattice of 16 tubes
connecting eight points on the rim to the vertices of an octagonal
frame, which is connected in turn to the mount structure. The dish
surface is thus suspended from the rim with no loading other than its
own weight. To prevent stresses from being built in to the dishes
during assembly, the fittings at the ends of the lattice tubes were
free to slide, and were then epoxied while the octagon frame was
suspended above the back surface of the dish on a jig. The antennas
can track over $\pm6$~hours in hour angle (HA) and --15$\degr$ to
$+90\degr$ in declination ($\delta$), although they cannot point to
negative elevation. In normal use an elevation limit of $20\degr$
restricts the HA range to $\pm2.5$ hours at $\delta$~=~--$10\degr$.

Each antenna is designed to be able to observe in up to $40\,$knots
($75\,\mathrm{km\,hr}^{-1}$) of wind, with an rms pointing accuracy of
$20\arcsec$, and survive up to $100\,$knots
($180\,\mathrm{km\,hr}^{-1}$). The antennas are housed in a reflective
aluminium enclosure (with internal dimensions of 40~m~$\times$~28~m at
ground level and height 4.5~m) to reduce ground spill and terrestrial
interference, and to offer additional protection from wind.

\subsubsection{Optics}
\label{sec:ant:sa:optics}

The aperture efficiency of classical Cassegrain systems is only in the
range of 70--75 per cent \citep{galindo64}. Shaped primary reflectors
are expensive to produce but have much better efficiency; for example
the efficiency of the LA calculated at 15~GHz is 89 per cent
(excluding the struts that support the secondary reflector; see
below). Our principal requirements were an efficiency of over 80 per
cent over most of the band with low sidelobes.

A rolled-edge Cassegrain system was developed to achieve these
criteria \citep{holler2007b}. An oversized secondary reflector
improves the aperture efficiency but also increases the sidelobe
level. These sidelobes can pick up emission from the warm ground and
degrade the system temperature, as well as introducing spurious
signals. Rolling the secondary reflector's edge towards the primary
reflector ensures that rays entering the horn come from the cold sky
rather than from the ground. Unlike a shaped Cassegrain system, the
phase-front in the rolled-edge Cassegrain system is not uniform, but
the distortion is negligible at 15~GHz. The residual sidelobes add
about 6~K to the system temperature.

An additional 15-cm rim around the primary reflector, mounted at
45$\degr$ to the optical axis, redirects part of the sidelobes to the
sky and reduces the ground spill to below 2~K. The aperture efficiency
is 81--82 per cent between 12.5~GHz and 15~GHz, falling to 75 per cent
at 17.5~GHz. Sidelobes at $>30\degr$ are suppressed by at least 60~dB.

The secondary reflector is supported by three struts; scattering from
the struts is reduced by shaped deflectors fitted to the struts to
redirect rays to the sky, giving a further reduction in ground spill.

\begin{figure}
\begin{center}
\includegraphics[width=7cm,origin=br,angle=0]{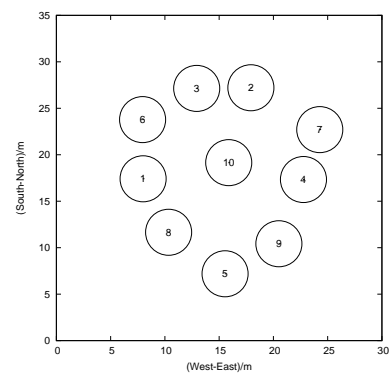}
\end{center}
\caption{Current SA relative antenna positions in
metres.\label{fig:sa-geom}}
\end{figure}

The SA is configured along a quasi-Reuleaux triangle \citep{keto1997}
along the north-south direction with a central antenna (Figure
\ref{fig:sa-geom}). Empirically, this offered the best compromise
between temperature sensitivity, coverage in the $uv$-plane (which
dictates the synthesised beam shape) and minimised antenna-to-antenna
shadowing within the confines of the enclosure. However, for dedicated
low-declination observations, stretching the array along the
north-south direction significantly improves the $uv$-coverage.

\begin{figure}
\begin{center}
\includegraphics[width=7cm,origin=br,angle=0]{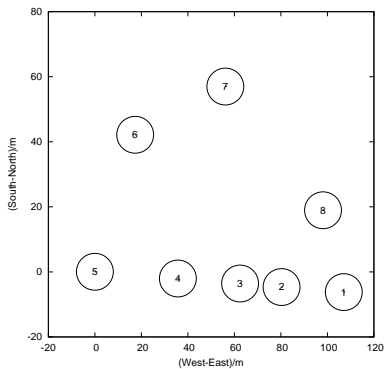}
\end{center}
\caption{LA relative antenna positions in metres.\label{fig:la-geom}}
\end{figure}

\subsubsection{Alignment}
\label{sec:ant:sa:align}

A north-south line was generated, and marked on brass plates on the
floor of the enclosure, from observations of Polaris in 2005
March. The SA antennas were aligned to this line with a Hilger and
Watts Microptic Theodolite Number 2 and levelled with a Taylor Hobson
digital clinometer. The antennas were aligned to within $1\arcmin$,
with further corrections made astronomically.

The antennas can be repositioned \eg to optimize $uv$-coverage for a
particular target or field. To facilitate this, we have installed an
optical alignment system that allows each antenna to be realigned
rapidly to within a few arcsec of its original orientation using a
mirror fixed to its structure. The system employs a Taylor Hobson
Micro-Alignment Telescope, operating in auto-collimation mode, in
conjunction with a moveable pentaprism.

\subsection{Large Array (LA)}
\label{sec:ant:la}

The 5-km Telescope was an approximately east-west linear array of
eight 13-m diameter Cassegrain antennas operating at 2.7, 5 and
15~GHz, with $\le$ 20~MHz bandwidth, to achieve a maximum resolution
of $0\farcs7$; it was designed for state-of-art imaging of radio
galaxies, radio quasars and supernova remnants. In 1985, work began on
equipping it with cryogenically cooled receivers and 0.35-GHz
bandwidth (with much narrower individual channels to avoid chromatic
aberration). This created the RT in order to do, among other things,
SZ imaging with its five closely-packable antennas. We have now moved
all eight antennas into a close, two-dimensional array and have fitted
new receivers and signal processing. The whole process is an example
of the continual reuse of expensive hardware over 40 years to keep
delivering new science.

\subsubsection{Array Configuration}
\label{sec:ant:la:config}

The LA antenna configuration is shown in Figure
\ref{fig:la-geom}. Antennas 1--4 are moveable along a railway track
and the positions of the remaining antennas are fixed. Antennas 6--8,
which were the most westerly of the 5-km Telescope, have been placed
to allow the array to observe equatorial fields without a loss of
resolution.

\section{Cryogenic Receivers}
\label{sec:cryo}

The first-stage amplifiers (section \ref{sec:cryo:amp}) are identical
for the SA and LA, but the cryostats and horns are very different
(sections \ref{sec:cryo:sa} and \ref{sec:cryo:la}).

\subsection{Low-Noise Amplifiers}
\label{sec:cryo:amp}

The first-stage amplifiers are a 8--18-GHz (optimized for 12--18~GHz
as standard) design by Marian Pospieszalski of NRAO Charlottesville,
and used as the IF amplifier in the NRAO 80--95-GHz VLBA receivers and
elsewhere. Each amplifier contains three InP high-electron-mobility
transistors (HEMTs). The amplifiers are intrinsically low-noise, with
a minimum noise temperature of about 6~K when cooled to 20~K.

The second-stage amplifiers are either earlier generation NRAO HEMT
designs (recycled from the CAT and RT) or internally-biased amplifiers
manufactured by Avantek; all are operated at ambient temperature. We
can afford to run the second-stage amplifiers warm because the gain of
the front-end amplifiers is so high (see figure
\ref{fig:if-system-la}). The bias of each front-end HEMT is externally
tunable from a `bias box'.

The noise temperature of the entire cryostat is 13~K at 15~GHz, rising
to 24~K at 12 and 18~GHz. The full system temperature, averaged over
12--18~GHz, is better than 30~K at zenith including the CMB and
atmosphere.

\subsection{SA Feed Assembly}
\label{sec:cryo:sa}

\begin{figure*}
\centering
\includegraphics[width=18cm]{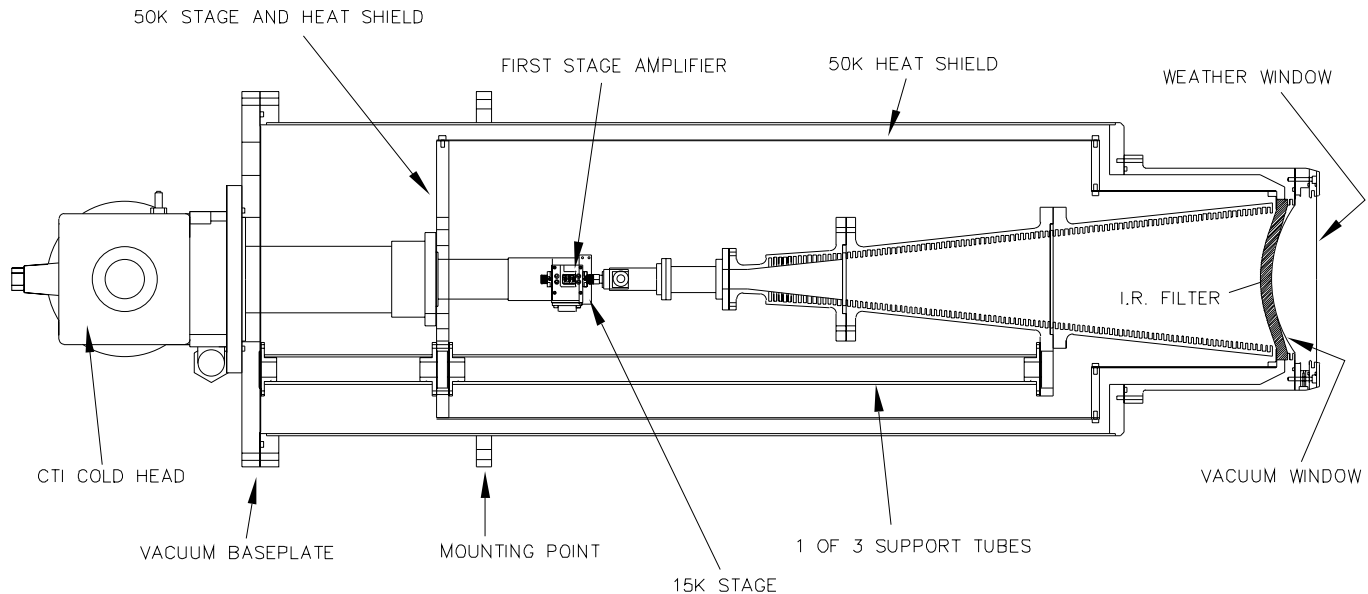}
\caption{SA cryogenic receiver. The internal diameter of the horn
mouth is 108 mm.}\label{fig:sa-cryostat}
\end{figure*}

The radio frequency (RF) signal from the secondary reflector enters a
corrugated feedhorn. Corrugated feedhorns have a circularly symmetric
aperture field distribution and low sidelobes (see \eg
\citealt{clarricoats73}). From the neck of the feedhorn, rectangular
waveguide is coupled to coaxial cable via a Flann Microwave adaptor,
resulting in sensitivity to Stokes' parameters I+Q. A cooled isolator
before the front-end amplifier ensures a good impedance match to the
amplifier and reduces the return loss. The feedhorn must be cooled to
20~K to achieve low receiver noise. Incorporating the entire feedhorn
into the cryostat and mounting everything on the vacuum base plate
simplifies the assembly. The feedhorn and the first-stage receiver are
cooled to 20~K by a closed-cycle helium refrigerator with a 50-K
intermediate shield surrounding it (Figure \ref{fig:sa-cryostat}). For
the SA we use CTI type 350C two-stage coldheads with cooling powers of
4 W at 21~K and 15 W at 77~K.

The disadvantage of incorporating the feedhorn in the cryostat is the
large (120-mm diameter) cryostat window. The window must be
transparent over the observational band and must maintain a good
vacuum seal under loading by atmospheric pressure, and an infra-red
filter is needed. Two designs were tried (see \citealt{kaneko2005},
\citealt{kaneko2006}). A 75-$\mu$m mylar weather window (and UV
filter) protects the vacuum window from the elements. Distributed dry
air is circulated between the two windows in order to prevent water
forming on the outside of the vacuum window.

We further found that dew formed on the cryostat weather window during
damp nights, typically raising the system temperature by a factor of
five. Dew formation is encouraged by the close proximity (see Figure
\ref{fig:sa-cryostat}) of the cooled feedhorn. So to keep the window
dry, a second dry-air system directs pressurized dry air over the
outer surface of the weather window. This solves the dew problem, as
well as drying the windows more quickly after rainfall.

\subsection{LA Feed Assembly}
\label{sec:cryo:la}

\begin{figure*}
\centering
\includegraphics[width=18cm]{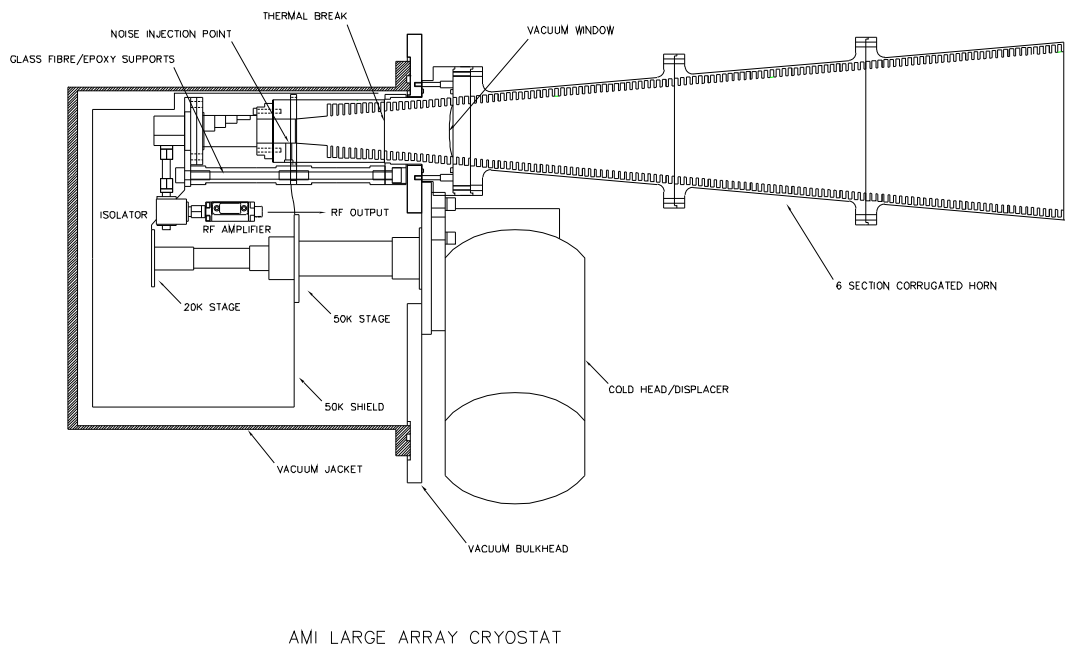}
\caption{LA cryogenic receiver. The internal diameter of the horn mouth is 106 mm.}\label{fig:la-cryostat}
\end{figure*}

The LA feed assembly (Figure \ref{fig:la-cryostat}) is based on the RT
15-GHz system. In contrast to the SA feed assembly, the horn is not
completely cooled. But the vacuum window aperture is consequently much
smaller and there is no need for an infra-red filter. There is a
thermal break in the horn between the throat and vacuum window. To
suppress undesirable mode effects, which were detected as peaks in the
cryostat system noise as a function of frequency, the first horn
section was fitted with a PTFE dielectric ring, secured in place in
the last corrugation closest to the throat. The SA and LA horns are
almost identical in size. As with the SA, the cryostat is cooled to
20~K in the first stage and 50~K in the second, but the LA has CTI
type 22C two-stage coldheads with cooling powers of 1.5 W at 21~K and
4 W at 63~K.

\section{Intermediate Frequency (IF) System}
\label{sec:if}

\begin{figure*}
\begin{center}
\includegraphics[width=22.5cm,origin=br,angle=90]{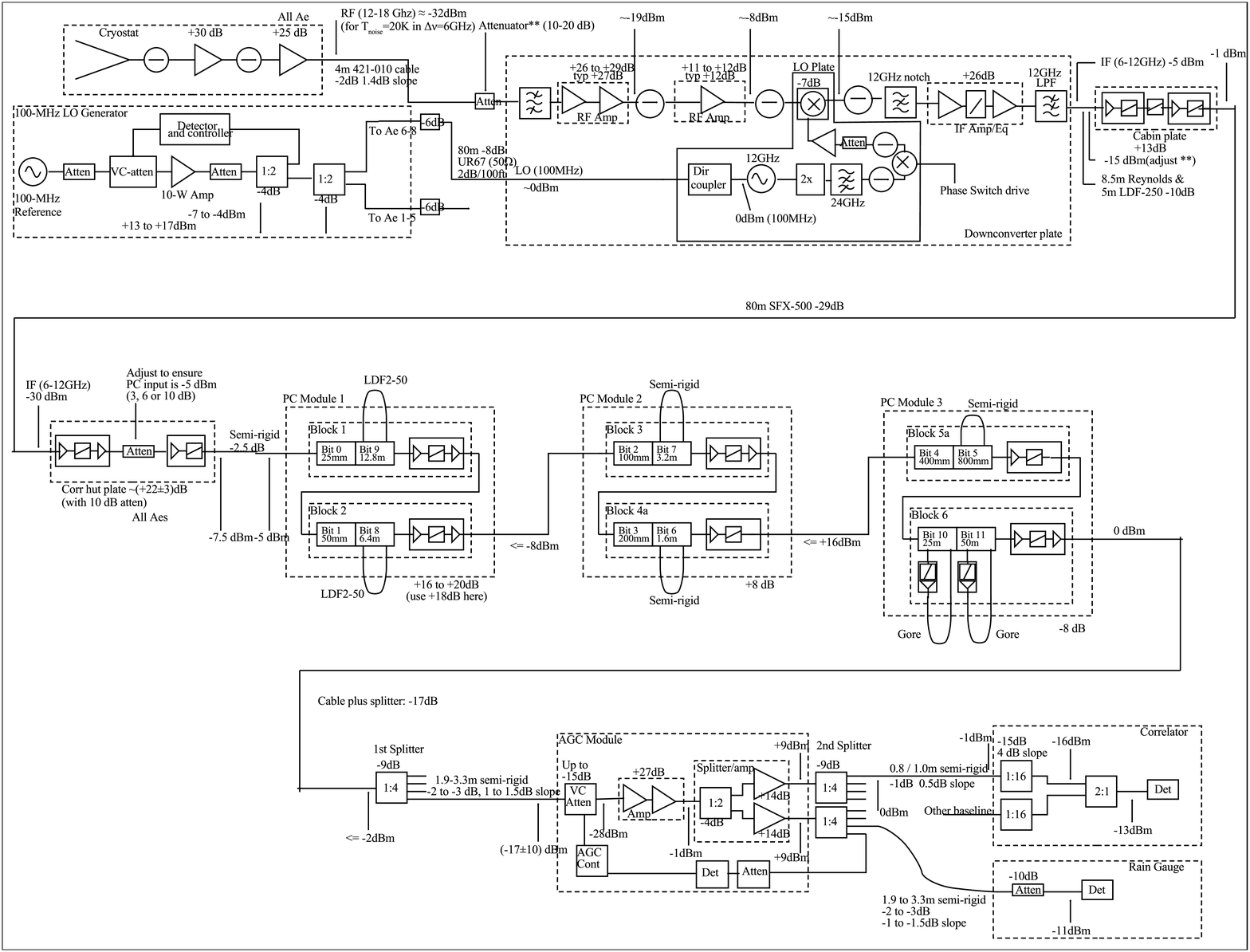}
\end{center}
\caption{IF chain for the LA. The SA IF chain is very similar, but
with a different LO distribution and without the amplifier-equalizer
plates either side of the long cable runs, and without the longest two
PC bits (see text for details). Power levels are in dBm where
0~dBm~=~1~mW. \label{fig:if-system-la}}
\end{figure*}

AMI's RF/IF chain is illustrated in Figure \ref{fig:if-system-la}. The
design is largely common to both arrays. The main differences are
outlined in section \ref{sec:if:differences}.

After the RF signal leaves the cryostat, it is downconverted (section
\ref{sec:if:dc}) using a local oscillator (LO, section
\ref{sec:if:lo}), to form the IF band. The total path lengths from the
horn to the correlator detector are kept as similar as possible
between, and within, antennas throughout the length of the IF
chain. The IF amplifier and equalizer units were designed and built
in-house, using off-the-shelf amplifier chips bonded to microstrip,
because the required performance was not commercially available. The
IF passes through the path compensators (PCs, section
\ref{sec:if:pcs}) and automatic gain control (AGC, section
\ref{sec:if:agc}) before being correlated (section \ref{sec:corr}).

\subsection{Local Oscillator}
\label{sec:if:lo}

In both arrays, a 12-GHz dielectric resonance oscillator is
phase-locked to a 100-MHz reference generated by a crystal
oscillator. The LO is then passed through a frequency doubler, and the
output is high-pass filtered to reject the original 12-GHz
signal. This 24-GHz LO is phase-switched (section
\ref{sec:if:dc:walsh}) and mixed with the 12--18-GHz RF to yield a
single-sideband IF of 6--12~GHz (see section \ref{sec:if:dc}). The
upper sideband is discarded. The lower sideband (with LO at 24~GHz) is
used in preference to the upper sideband (LO at 6 GHz) in order to
keep the LO and its harmonics outside the IF band. A fundamental
oscillator at 24~GHz would have been preferred to the doubled 12-GHz
system but no suitable device was available.

In the LA, the long LO distribution paths mean that cable losses would
be too great to distribute the LO at 12~GHz. Instead, the 100-MHz
locking signal is distributed, via UR67 coaxial cable, and each
antenna has its own 12-GHz phase-locked oscillator and 24-GHz
doubler. Each LA downconverter originally had its own AGC to control
the 100-MHz power level. Small variations in 100-MHz power were found
to have minimal effect on the 24-GHz stability, and the
voltage-controlled attenuator was found to be noisy and unstable, so
the AGC was removed.

In the SA, the LO distribution paths are shorter and the corresponding
12-GHz cable losses less, so that it is more economical to have a
single, central 12-GHz phase-locked oscillator locked to a 100-MHz
oscillator, distributed at 12~GHz via Andrews LDF2-50 coaxial cable,
with doublers and filters at each antenna.

Bench tests indicated that the LO multiplier units at each LA antenna
worked adequately, but when installed they locked only intermittently
to the 100-MHz phase-reference signal; this was due to their being
operated at close to the maximum temperature in their specification,
necessitating re-engineering of the downconversion temperature
control.

\subsection{Downconversion}
\label{sec:if:dc}

The RF is sent from the cryostat to the downconversion system via a
coaxial cable with microporous teflon dielectric (Reynolds 421-010 on
the SA and Tensolite 301 on the LA). The downconversion system, which
includes modules by Atlantic Microwave Ltd, is mounted on a
temperature-controlled plate kept to within $pm3\,\degr$C of
$35\degr$C in order to reduce temperature-dependent path changes, and,
more importantly, to ensure the performance stability of the active
components. We undertake downconversion partly because the manufacture
of amplifiers, switches and other components is much more economical
at lower frequencies. In such a heterodyne system, the instrument's
observing frequency can also later be adjusted by adjusting the
frequency of the LO.

A bandpass filter on the downconverter plate restricts the bandpass to
12--18~GHz and the RF then passes through two amplifiers and
isolators. A 12-GHz notch filter removes any stray LO
contamination. Another filter rejects geosynchronous satellite signals
around 12.8~GHz.

The IF is transmitted to the correlator room using a 55-m length of
Andrews' LDF2-50 cable on the SA and an 80-m length of SFX-500 cable
on the LA. At either end of the cable the signal is amplified and
equalized to remove the frequency slope due to the cables' increasing
loss with frequency.

\subsubsection{Phase Switching}
\label{sec:if:dc:walsh}

Partly to attenuate spurious signals and the effects of drifts in the
signal processing, we introduce 180$\degr$ phase-switching early on in
the system (by switching the 24-GHz LO with a known rectilinear
waveform) and synchronously detect this switching late in the system
after correlation (just before the correlator readout). This process
produces the modulation in power of the product signal for each
baseline and allow the use of a power detector rather than a
multiplier in the correlator \citep{ryle52}.

The waveforms used for phase-switching are Walsh functions (see \eg
\citealt{walsh}). Their key property is that they are orthogonal,
independent of relative phase. We use a different odd-numbered
function for each antenna in order to remove the effect of any signal
leakage from one radiometer path to another. Walsh functions also make
it easy to construct the particular synchronous detection function to
apply to each correlation.

The LO is modulated by using a balanced mixer with a DC-coupled IF
port --- this allows the LO phase to be controlled by injecting a
positive or negative current in to this port. A feedback system (see
Section \ref{sec:corr:pswb}) controls the magnitude of these bias
currents such that the amplitide of the LO is equal in each phase
state.

We estimate that our phase-switching process reduces spurious signals
and drifts by 30 dB.

\subsection{Path Compensators}
\label{sec:if:pcs}

Radiation received from a source on the sky takes longer to reach one
antenna of a baseline than the other, since one is further from the
source. We must continually switch in extra fixed electrical lengths
into every antenna IF, thus ensuring that the difference in electrical
length between the sky point and correlator by any path is much less
than the coherence length of the radiation,
$c/\Delta\nu_{\mathrm{ch}}$, where $\Delta\nu_{\mathrm{ch}}$ is the
bandwidth of a frequency channel. This is the function of the path
compensators (PCs). Note that because this path compensation is
continual rather than continuous, the phase response of an
interferometer to the sky point changes at the fringe rate as the
earth rotates. Note also that because this path compensation is
introduced in the IF whereas the initial path error is in the RF, the
phase of the visibility changes at a rate proportional to the change
of path length times the difference between the RF and IF frequencies.

On entry to the correlator room of each array, the IF for each antenna
passes to the PC modules, where fixed paths of different electrical
lengths are switched in.

There are 10 PC bits for each antenna of the SA, with lengths from 25
mm up to 12.8 m (see also section \ref{sec:if:differences}). Each bit
is nominally slightly less than twice as long as the previous one to
ensure that path compensation can be provided for every reachable part
of the sky.

The PCs were manufactured in-house as such units are not commercially
available. RF switches on a microstrip substrate, controlled by the
correlator microprocessor (section \ref{sec:control}), select between
a shorter and a longer path for each bit. The equalization and gain of
the shorter path are adjusted by placing a piece of Eccosorb above the
path's microstrip. In this way, the gain and equalization of the
shorter path can be matched to that of the longer to within $\approx
0.5$dB. In order of increasing length, the switched bits are formed
from microstrip, aluminium or copper semi-rigid coaxial cable, Andrews
LDF2-50 coaxial cable and, for the longest two bits of the LA only,
Gore PRP030306-123 coaxial cable.

\subsection{Automatic Gain Control}
\label{sec:if:agc}

After the PCs, the IF is split passively and equally using Wilkinson
dividers. Following the first split, an aluminium semi-rigid cable
takes the IF to an automatic gain control (AGC) module. The AGC
ensures that, despite the time-varying and intrinsically different
gains of components, as well as varying atmospheric conditions, the IF
power entering the correlator is constant to within $\approx 1$dB (in
part to ensure that each correlator diode operates efficiently). The
AGCs handle an input range of --23~dBm to --3~dBm.

After the AGC, the IF is split again, with one of the splitter outputs
returning for the AGC detector loop. The IFs from each pair of
antennas then go to two correlator boards (section \ref{sec:corr}).

The output of the AGC is sensed by a detector, which feeds back to a
Hittite HMC 121 voltage-controlled attenuator. This has better
stability, and phase and frequency performance, than a variable-gain
amplifier and gives up to 30~dB of control over the IF band. An
amplifier is needed to compensate for the attenuator loss. The
feedback detector is a coaxial Schottky diode detector (from Advanced
Control Components), with a square-law characteristic over the power
control range. The detected voltage provides negative feedback control
to the voltage-controlled attenuator.

\subsection{Differences between the SA and LA}
\label{sec:if:differences}

The IF chains for the two arrays are almost identical. The LA has some
extra components which are summarized here:

\begin{itemize}

\item Two more path-compensator bits, with cables of 25 and 50 metres
to allow for the much longer baselines; we eventually sourced a
particular cable type (PRP030306-123) manufactured by Gore to give
low-enough dispersion;

\item The LO is distributed at 100~MHz and multiplied at each
antenna's downconverter, rather than being distributed at 12~GHz (see
section \ref{sec:if:lo});

\item As the IF cables from each antenna are much longer, extra
amplifiers and equalizers are needed at both ends of the long cable
runs to compensate for the extra loss and slope caused by these
cables.

\end{itemize}

\section{Correlator}
\label{sec:corr}

\subsection{RF Design}
\label{sec:corr:design}

The AMI correlator is an analogue `add and square' Fourier Transform
Spectrometer (FTS) (for full details see \citealt{holler2007a}). It
splits the 6~GHz band into eight complex frequency channels of
0.75~GHz by correlating the signals of each baseline simultaneously at
16 different discrete delays and then Fourier transforming the
measurements in software (a so-called XF correlator, i.e. first
correlate, then Fourier transform). The discrete delay length of the
lags in the FTS is set by the design bandwidth of 6~GHz and has to
satisfy Nyquist's sampling theorem, resulting in a nominal length of
25~mm for each lag.

In general there are two possible architectures for a lag correlator,
`complex' and `real'. The complex correlator measures amplitude and
phase of the detected IF signal at each correlation, using the Nyquist
sampling rate for the delay length. The imaginary part is produced by
a broadband $90^{\circ}$ phase shift in one antenna. The real
correlator measures amplitude only, with half the nominal delay
spacing, resulting in the same total number of correlations.

In the `add and square' design of the FTS the number of detectors has
to be doubled in order to reach the same signal-to-noise performance
as with a multiplying correlator. For each baseline the sum (`+') and
difference (`--') of the two inputs are formed, the latter using an
additional broadband $180^{\circ}$ phase shift in one antenna, and
their powers detected with a diode detector. This results in two
independent measurements for each lag and the signal-to-noise is
increased by $\sqrt 2$ over a single + or -- channel. Previous
implementations of similar systems, \eg the VSA (see
\citeauthor{rusholme2001}), combined the + and -- correlations in
hardware. However, in order to be able to calibrate each for hardware
systematics (especially unequal lag spacing) separately, we postpone
the combination of + and -- signals to the mapping process.

\begin{figure}
\begin{center}
\includegraphics[width=8cm,origin=br,angle=0]{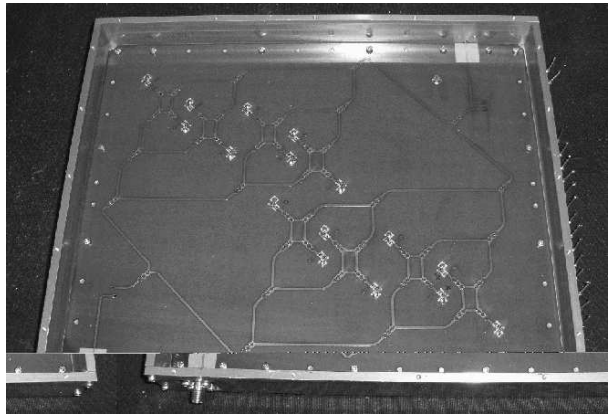}
\end{center}
\caption{Correlator microstrip. The IFs from the two antennas enter
bottom left and top right, before fanning out into a tree of lags. The
detectors are at the tips of the branches.\label{fig:corr-microstrip}}
\end{figure}

The lags are implemented using microstrip delay lines (Figure
\ref{fig:corr-microstrip}). Fluctuations in the dielectric constant
of the substrate produces irregularities in the lag spacing, giving an
imbalance between the real and imaginary correlated components for
each baseline; there are also different intrinsic gains in the
detectors and dispersion over the wide band. This leads to a unique
frequency response for each correlator, since the mean lag spacing is
slightly different. Further, the correlator lags are systematically
slightly long, which leads to aliasing in the first channel.

The function of each correlator detector is to rectify the IF signal,
with output proportional to each individual rectified peak,
independently of amplitude. To do this, the rectifier output must see
a following impedance that is high and purely
resistive. Unfortunately, the final design resulted in the diode
seeing a largely capacitive load. Even the best lumped components have
resonant frequencies of $\sim$~10~GHz, but the use of a series
resistor after the diode and the keeping of distance to the input of
the following op-amp as short as possible would have given better
performance. As it is, this difficulty has led to substantial loss of
sensitivity for the top two IF channels, i.e. the bottom two RF
channels. However, due to problems with geostationary satellite
emission, these two channels are currently suppressed anyway by
hardware filters. Further, the correlator lags are systematically too
long, which leads to aliasing in the first channel.

The SA and LA correlators are identical, but separate. Since each
correlator module implements either the + or the -- correlation for
one baseline the SA has 90 correlators for its 10 antennas, and the LA
has 56 correlators for its eight antennas.

\subsection{Synchronous Monitoring of System Temperature}
\label{sec:corr:rain}

The relative system temperature of each antenna is monitored by a
`rain gauge', which detects the power received from
amplitude-modulated noise injected by a probe in the
circular-waveguide throat of the receiver horn. The increase in the
system temperature is negligible from the point of view of overall
noise performance, but the injected signal is easily detected by the
synchronous power detector. As the system temperature fluctuates due
to \eg gain variations or changing atmospheric emissivity, the AGC
adjusts the IF signal amplitude and this decreases or increases the
relative amplitude of the modulated noise. The rain gauge measures
this and hence the system temperature. All the noise sources are
switched with the same Walsh function, which is not one of those used
for phase switching.

\subsection{Phase-switch Balance}
\label{sec:corr:pswb}

A phase-switch balance feedback loop uses a similar method to correct
for any amplitude modulation caused by the LO power being different in
the two states of the phase-switch mixer (see section
\ref{sec:if:dc:walsh}). A synchronous detector generates a feedback
signal that is used to bias the mixer control voltages and hence
eliminate any amplitude modulation.

\subsection{Correlator Readout}
\label{sec:corr:readout}

The readout system is described in \citet{kaneko2005} and
\citet{holler2007a}. For each baseline, each pair of correlators is
served by two readout cards. The phase-switching signals introduced at
downconversion are demodulated and integrated in an analogue circuit,
and the signal can then be digitized at a relatively modest rate. We
use a 20-bit analogue-to-digital converter to ensure that there is
sufficient dynamic range and that the quantization noise is much less
than the system noise. Correlators are read out in `racks', each of
these consisting of three sets of correlator-readout systems. Each
rack interfaces serially to the correlator microprocessor (see section
\ref{sec:control:data}).

\section{Control System}
\label{sec:control}

The control systems for the SA and LA are independent realizations of
the arrangement illustrated in Figure \ref{fig:sa-system}. At the
heart of each is a local Ethernet segment connecting a number of
rack-mounted, single-board microprocessors that control the drive and
monitoring processes and handle the data stream from the correlator
readout. In the case of the SA, these are Eurotech microprocessors
(\texttt{www.arcom.com}) running the QNX operating system
(\texttt{www.qnx.com}), with a standard desktop computer acting as the
master node; the LA control system is a development of this using a
later model of Eurotech microprocessors, which run a variant of Linux
rather than QNX. In both cases, the main telescope control system is a
Sun workstation which is connected to both the local and wider
Ethernets; this runs the user interface, synchronizes telescope
control with data readout and deals with the archiving of the raw
time-stream data to disk.

\subsection{Data Acquisition}
\label{sec:control:data}

Data are transmitted as serial digital signals from the correlator
racks to the correlator microprocessor, which in the case of the SA is
a slave node of the QNX network. Racks are identified by header and
footer numbers, which also serve as error checks.

The correlator microprocessor samples data at 32~Hz and passes it via
ethernet to a shared memory segment in the Sun workstation, which
waits for input.

The only processing steps performed on the data in real time are to
remove path-compensator switching transients, by flagging the first
two 32-Hz samples when a PC setting changes, and to average samples
(over one second for the SA and 1/4 second for the LA) before writing
to disk in a local data format. Raw data rates from the correlator are
of the order of 200~KBps. The data format consists of a header that
describes the observation and the telescope configuration, followed by
the time-stream data for all active baselines and monitoring
information. Each night, the telescope data files are transferred from
the observatory to an archive at the Cavendish Laboratory via an
optical-fibre data link.

The correlator microprocessor also sets the state of the path
compensator bits, in response to on-the-fly requests from the drive
microprocessor. It further reads the rain-gauge data (section
\ref{sec:corr:rain}) via similar analogue-to-digital converters to
those for the correlator data.

\subsection{Drive Control}
\label{sec:control:drive}

\subsubsection{Small Array}
\label{sec:control:drive:sa}

In the SA, a single microprocessor controls the drives for all of the
antennas in the array, via a Profibus serial field bus
(\texttt{www.profibus.com/pb}). Each antenna has HA and $\delta$
Eurotherm servo amplifiers (\texttt{www.eurotherm.co.uk}) and two
Heidenhain encoders (one for each axis, \texttt{www.heidenhain.com}),
which are devices on the serial bus. The encoders provide a feedback
loop for positioning each drive. The logic for limit switches and
emergency stop interlocks is handled by relays at each antenna.

Every second, the drive microprocessor fetches the encoder readings
(that is, current positions) of all 20 drives from the encoder
gateways, determines the speed required to reach the requested encoder
positions received from the drive control process in the Sun
workstation, and passes these speeds to the servo amplifiers, thus
completing the feedback loop. The encoder readings are passed back to
the Sun workstation for report and error handling.

Each antenna uses a pair of shielded, dual-core cables to the Profibus
standard. The serial bus is completed in the correlator room, to allow
each antenna to have a dedicated bundle (that includes power, LO, IF,
dry air and monitoring cables) for easy repositioning, rather than
daisy-chaining the antennas directly to each other. An electronic
repeater is placed in the correlator room after antenna 5. Active
termination of the Profibus was not necessary.

A HA--$\delta$ combined south limit switch allows the drives to reach
a lower declination than with a single, hard south limit. The
elevation limit is +20$\degr$. The SA antenna slewing speed is
$48\degr\mathrm{min}^{-1}$ in HA and $13\degr\mathrm{min}^{-1}$ in
$\delta$.

\subsubsection{Large Array}
\label{sec:control:drive:la}

Drive control for the LA is similar to that for the SA, except that
the LA has one Eurotech drive microprocessor per antenna. Rather than
using Profibus, each microprocessor communicates directly with the
Eurotherm motor controllers of its respective antenna via an analogue
connection.

The other difference is that for the LA, the drive control feedback
loop is completed in the Sun workstation. There is a master
drive-control task that performs observation control and computes
target pointing values for all antennas, and separate independent
communications processes for each antenna that compare the actual and
required encoder readings each second, compute the appropriate speed
commands and transmit these to each microprocessor using a separate
Internet-socket connection for each microprocessor. Communication
among the control processes in the Sun workstation is via data
structures in the shared memory segment. The LA antenna slewing speed
is $15\degr\mathrm{min}^{-1}$ in each axis and the elevation limit is
+5$\degr$.

\subsection{Monitoring}
\label{sec:control:monitor}

The cryostat temperature and pressure, the front-end amplifier biases,
interlock states and the AGC feedback voltages are relayed to the
monitor microprocessor once every second. We also monitor the
temperatures of the downconversion plates and the state of the PLO
lock. The wind speed is monitored continuously and the LA stowed if
the speed reaches a predetermined level.

\section{Lessons Learned}
\label{sec:comm}

In the course of construction, various problems were of course found,
such as the need for extra air blowers on the SA (see section
\ref{sec:cryo:sa}), the LO locking (section \ref{sec:if:lo}), the
correlator response, unequally spaced and dispersive lags
(\ref{sec:corr:design}). There are three other noteworthy issues:

\begin{enumerate}

\item In this and other telescopes that we have constructed, we have
found that often for very simple `components' the manufacturers'
specifications were inadequate to predict actual performance. An
example that caused problems in the current context is the operating
temperature range of drive-gearbox grease.

\item Both real and complex correlators should have identical
sensitivities (in the limit of a large number of frequency
channels). We prototyped both versions and opted for the real
correlator since the RF board was slightly smaller and to obviate
small phase errors introduced by the $90^{\circ}$ phase
shifter. However, this means that full phase information is not
available until after the Fourier transform is performed in
software. At several points during commissioning having phase
information at an earlier stage would have been useful. We advise
others considering building this sort of correlator to consider
carefully whether to build a real or complex correlator.

\item We found that the LDF2-50 cable used on the SA for the IF and
longer PC cables had significant dispersion over the AMI band. In the
IF, as long as the dispersion of each cable is the same, this did not
matter. However, in the PC, in which air path to one telescope is
compensated by electrical path in the other, dispersive cable is a
problem. We calculated that for PC total lengths in which the longest
bit is 12~m, the use of LDF2-50 was acceptable and could be corrected
for in software. But on the LA, where PC cables extend to 50~m, cable
of much lower dispersion is essential but proved very difficult to
find.

\end{enumerate}

\section{Performance}
\label{sec:concl}

Both the SA and the LA have been commissioned. Both arrays meet their
design sensitivity, with lower system temperatures compensating for
the effect of the loss of the bottom two frequency channels. In one
hour, the SA has a 1-$\sigma$ flux sensitivity of 500~$\mu$Jy and the
LA has a 1-$\sigma$ flux sensitivity of 50~$\mu$Jy.

\begin{figure*}
\centering
\mbox{
\subfigure[Single-pointing SA map.]{\label{fig:joint-sa}\includegraphics[width=8cm,origin=br,angle=0]{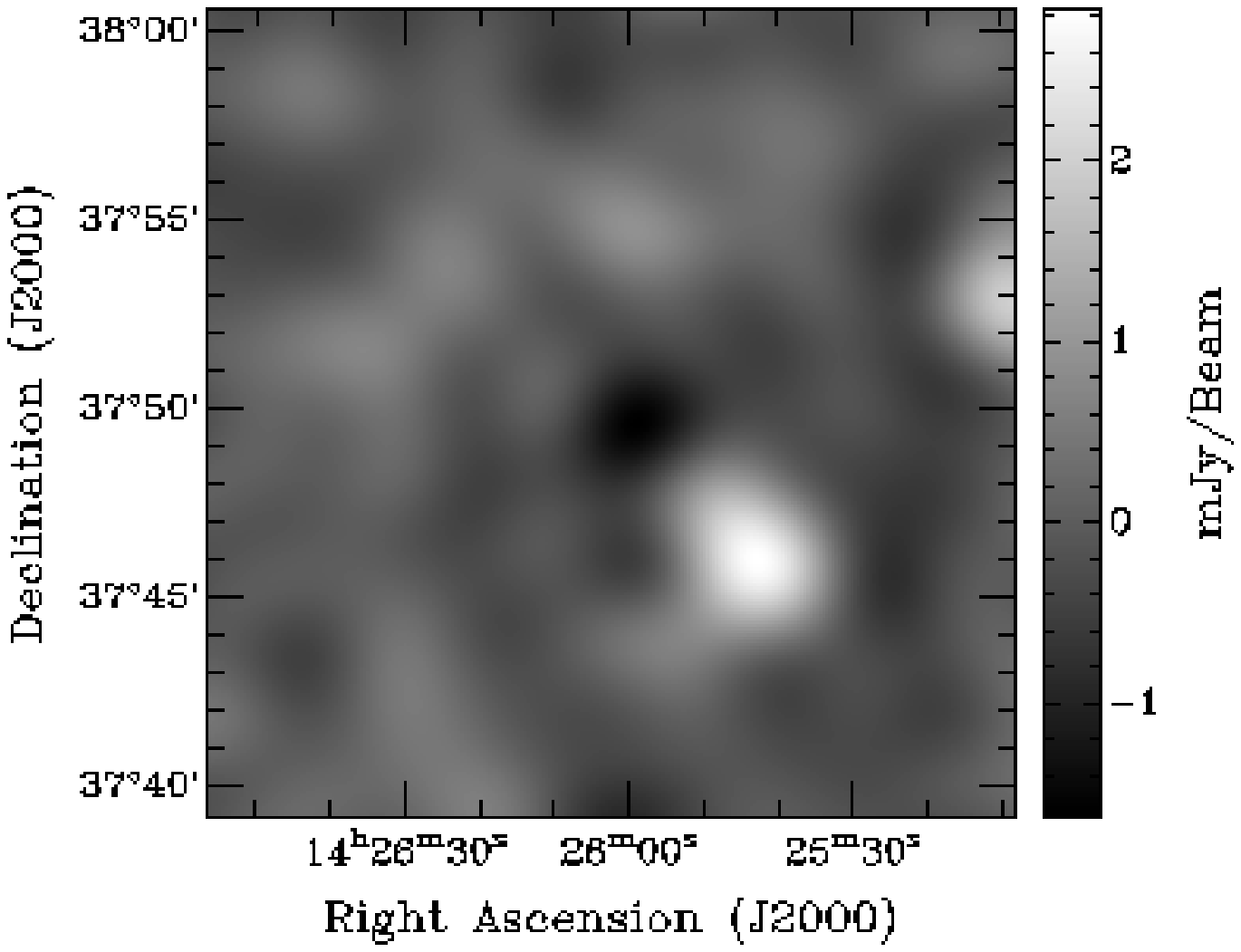}}}
\mbox{\subfigure[LA map comprising a 7- and 19-point raster.]{\label{fig:joint-la}\includegraphics[width=8cm,origin=br,angle=0]{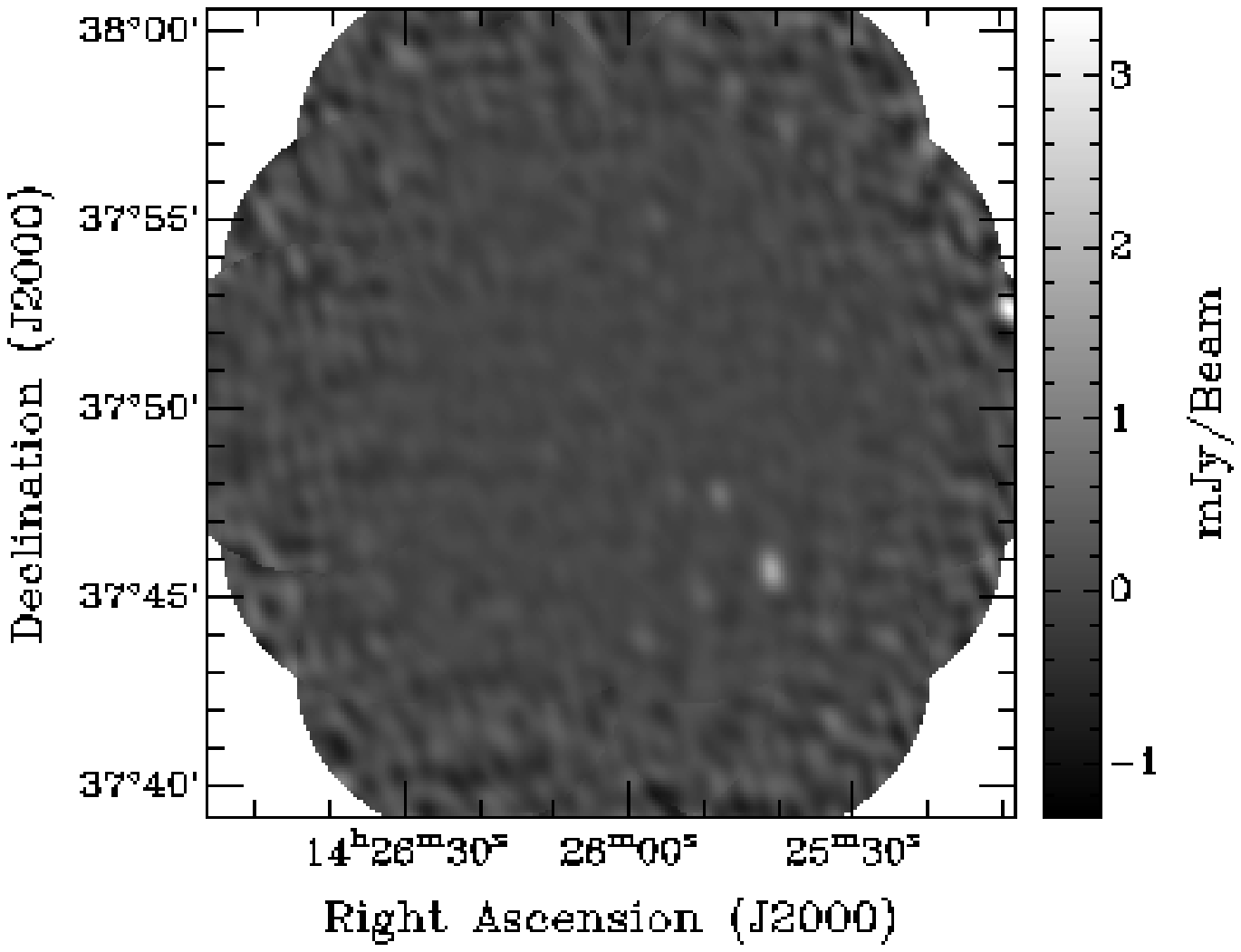}}}\quad \\
\caption{A simultaneous observation of Abell 1914 with AMI.\label{fig:joint}}
\end{figure*}

Figure \ref{fig:joint} shows a simultaneous observation with both
arrays of the cluster Abell 1914 (see \eg AMI Collaboration
\citeyear{ami-a1914}). The SZ effect from the cluster is clearly
visible at the centre of the SA map (Figure \ref{fig:joint-sa}). Both
the SA and LA maps show a number of radio point sources; these are
unresolved in the SA map, but separated in the LA map. The SA
observing time was 14 hours and the LA map comprises two 10-hour
rasters of seven and 19 points. The LA observations used just six of
the eight antennas, as two antennas were out of service at the time.

\section*{Acknowledgments}
\label{sec:acknowledgments}

We thank PPARC/STFC and Cambridge University for AMI construction and
operation. TC, MD, TF, NHW, TK, KL, PM, CRG, AS, TS and JZ acknowledge
PPARC/STFC studentships. FF acknowledges support from the Cambridge
Commonwealth Trust and the Cambridge Isaac Newton Trust, JG
acknowledges support from the Cambridge Isaac Newton Trust and the
Cambridge European Trust, JP acknowledges support from the Cambridge
Overseas Trust and MV acknowledges support from the Gates Cambridge
Trust. We thank Paul Alexander and Steve Rawlings for useful comments
on the paper. We thank NRAO Charlottesville for superb help with the
front-end amplifiers. We are also grateful to Atlantic Microwave Ltd,
Castle Microwave Ltd, W.~L.~Gore \& Associates Inc., Steven Hardwick
Associates Ltd and Prima Electronic Services Ltd for their excellent
technical understanding and assistance.

\bibliographystyle{mn2e}
\bibliography{ami-zero}

\bsp

\label{lastpage}


\end{document}